\documentclass[journal=langd5,manuscript=article]{achemso}
\usepackage{graphicx}
\usepackage[tight,centerlast]{subfigure}

\usepackage[version=3]{mhchem} 
\usepackage[T1]{fontenc}       

\author{Philipp Gutfreund}
\email{gutfreund@ill.eu}
\affiliation[ILL]{Institut Laue-Langevin, 38000 Grenoble, France}

\author{Marco Maccarini}
\affiliation[ILL]{Institut Laue-Langevin, 38000 Grenoble, France}
\alsoaffiliation[Universit\'e Joseph Fourier]{TIMC-IMAG, Universit\'e Joseph Fourier, 38706 Grenoble, France}

\author{Andrew J.C. Dennison}
\affiliation[TU Berlin]{Department of Chemistry, Technical University Berlin, 10623 Berlin, Germany}
\alsoaffiliation[ILL]{Institut Laue-Langevin, 38000 Grenoble, France}
\alsoaffiliation[Sheffield]{Department of Physics and Astronomy, University of Sheffield, S102TN Sheffield, UK}

\author{Max Wolff}
\affiliation[Uppsala University]{Materials Physics, Department of Physics and Astronomy, Uppsala University, 75121 Uppsala, Sweden}

\title[The search for Nanobubbles by using Neutron Reflectivity]{The search for Nanobubbles by using specular and off-specular Neutron Reflectometry}
\begin{document}

\begin{abstract}
We apply specular and off-specular neutron reflection at the hydrophobic silicon/water interface to check for evidence of nanoscopic air bubbles whose presence is claimed after an {\it ad hoc} procedure of solvent exchange. Nanobubbles and/or a depletion layer at the hydrophobic/water interface have long been discussed and generated a plethora of controversial scientific results. By combining neutron reflectometry (NR), off-specular reflectometry (OSS) and grazing incidence small angle neutron scattering (GISANS), we studied the interface between hydrophobized silicon and heavy water before and after saturation with nitrogen gas. Our  specular reflectometry results can be interpreted by assuming a sub-molecular sized depletion layer and the off-specular measurements show no change with nitrogen super saturated water. This picture is consistent with the assumption that, following the solvent exchange, no additional nanobubbles are introduced at significant concentrations (if present at all). Furthermore, we discuss the results, in terms of the maximum surface coverage of nanobubbles that could be present on the hydrophobic surface compatibly with the sensitivity limit of these techniques.
\end{abstract}

\section{Introduction}
The solid-water interface has been for many years the subject of fundamental scientific research due to the relevance of interfacial phenomena to biological processes in which water plays a major role \cite{Ball2008}. Moreover, the push for  miniaturization driven by rapidly evolving microfluidic technologies \cite{Thorsen2002} calls for a deeper understanding of the physical-chemical mechanisms that govern the processes occurring at this interface. Experiments using different model hydrophobic surfaces at the solid-water interface showed controversial results as a liquid layer of reduced density was observed sandwiched between the hydrophobic surface and the bulk water \cite{Maccarini2007}. In order to quantify the resulting depletion effects, the depletion distance $d_{2}$ \cite{Mamatkulov2004} was introduced:
\begin{equation}
	d_{2}=\int \left( 1-\frac{\rho(z)}{\rho_{bulk}} \right) \, dz ,
	\label{eq:D2}
\end{equation}
where $\rho(z)$ denotes the density of the depleted liquid at a distance \textit{z} from the interface and $\rho_{bulk}$ represents the bulk liquid density. $d_{2}$ reduces the smeared-out density profile of the depletion to a step-like function that represents an equivalent layer of zero density. The results of neutron reflectometry (NR) studies on spin-coated deuterated polystyrene ($d$PS)-D$_{2}$O interfaces showed a depletion distance of $d_{2}\approx2.6$\,\AA\ \cite{Steitz2003}, however, a repeat of this experiment by another group \cite{Seo2006} did not show this depletion  when a freshly prepared PS film was not exposed to air before the measurement. A similar scenario emerged with X-ray reflectometry (XRR) results on bulk water in contact with octadecyl-trichlorosilane (OTS) coated substrates \cite{Mezger2010}. A depletion distance of only $d_{2}=1.1$\,\AA\ was observed although up to half of the contribution may have arisen from the hydrogen termination of the hydrophobic coating, which is practically invisible to X-rays \cite{Ocko2008}. Even though this topic is still controversial, the depletion between water and a hydrophobic surface, if present at all, might occur on a sub-molecular scale and may arise from preferred orientation of water molecules at the interface as suggested by molecular dynamics (MD) simulations \cite{Mezger2010} and Raman-scattering \cite{Davis2012} and demonstrated for other liquids at hydrophobic surfaces \cite{Schmatko2005,MaccariniSteitz2007,Gutfreund2011,Gutfreund2013}.\par
The situation becomes increasingly complicated if water supersaturated with gas is used as the exceeding gas may condense in the form of a nanoscale layer at a hydrophobic interface \cite{Zhang2007}. These so-called nanobubbles were initially assumed to be the cause of the hydrophobic gap \cite{Steitz2003} and motivated further reflectometry studies to probe the influence of different gas enrichments on  the water depletion yielding controversial results \cite{Doshi2005,Mezger2006}. A further argument of controversy is the fact that theoretical calculations predicted the lifetime of nanobubbles to be in the $\mu$s range \cite{Attard2003,Brenner2008,Colaco2009} and thus not observable on a laboratory time scale. Therefore the observation of nanobubbles has often been attributed to the invasive effect of the Atomic Force Microscopy (AFM) technique \cite{Ishida2000,Tyrell2001}, which was mainly used to show their presence. With the introduction of a well established protocol to make nanobubbles reproducibly and at high surface coverage in this system \cite{Lou2001}, the so-called solvent exchange technique, it became possible to deliberately produce nanobubbles which are stable on the time scale of hours or even days \cite{Zhang2008} by replacing ethanol by water. Eventually, nanobubbles were observed with non-invasive techniques like infrared spectroscopy \cite{Zhang2007} and optical microscopy \cite{Karpitschka2012}. However, though it is currently accepted that nanobubbles may be artificially produced locally at solid-liquid interfaces under certain circumstances \cite{Peng2014} their (time and space averaged) concentration and thus their influence on mesoscopic quantities in microfluidics or cell biology is still debated \cite{Ball2012}. We believe this is due to the fact that the techniques used so far to examine nanobubbles are either local (various microscopy techniques) or they had no particular sensitivity to the shape of the bubbles, and quantifying only the mean surface coverage (infrared spectroscopy, specular reflectometry).\\
The reported size of nanobubbles (50 - 10000\,nm) makes them ideal candidates for investigations with grazing incidence small angle neutron scattering (GISANS), specular and off-specular neutron reflectometry. These techniques operate in the reciprocal space and provide structural information averaged in time and over a macroscopic surface area in contrast to the aforementioned local microscopy techniques \cite{Daillant}. The investigation of buried interfaces, e.g. solid-liquid interfaces, is mainly done by NR or high energy X-rays as those interfaces are rarely accessible by soft X-rays or visible light. Neutrons have a further advantage for soft matter interfaces since cold and thermal neutrons do not cause radiation damage of organic specimens nor they cause alterations of the samples due to radiation induced accumulation of charges on the surfaces \cite{Ghosh2013}. Neutron scattering techniques can  further augment their capabilities thanks to isotopic substitution, which offers a powerful tool for contrast enhancement of the low atomic number elements typically composing soft matter \cite{Squires}. Off-specular scattering (OSS), which is typically several orders of magnitude weaker than the specular reflection is not routinely used, apart from synchrotron X-rays sources \cite{Daillant}. However, specific examples where neutron OSS has been used are highly ordered systems like magnetic domains or gratings, especially in multilayered systems\cite{Zabel2007}, or for microphase separated polymers in heavy water \cite{Wolff2009}. Other examples include capillary waves\cite{Sferrazza1998}, lipid bilayers \cite{Salditt2005,Jablin2011}, polymer dewetting \cite{deSilva2009} and organic photovoltaics \cite{James2015}, which all profit from the high contrast of deuterated materials. Grazing incidence small angle neutron scattering has recently experienced an increasing recognition\cite{MuellerBuschbaum2015} partly due to the unique possibility to use TOF in combination with GISANS to measure depth dependent patterns in a single measurement \cite{Metwalli2011,Wolff2014} and due to advances in analysis software \cite{BornAgain}.\par
Previous GISANS measurements \cite{SteitzReport2004} on the interface between $d$PS and D$_{2}$O were performed on D22 at the Institut Laue-Langevin (ILL) in Grenoble, France, but the PS layer was unstable at higher temperatures used to enhance the appearance of nanobubbes. Therefore we used in this study silicon  substrates coated with OTS that are stable at the relevant range of temperature and performed specular and off-specular  NR and GISANS on N$_{2}$ enriched D$_{2}$O in order to cover the relevant lateral and out of plane length-scales of depletion layers and nanobubbles. 

\section{Experimental Details}
The silanization of the single crystal silicon (100) block (80*50*20 mm$^{3}$, Siltronix, France) was performed according to reference \cite{Wang2003}\,. The advancing contact angle of \textit{Millipore} filtered water was 102$^{\circ}$ and the receding one 70$^{\circ}$ (static contact angle was 93$^{\circ}$). The specular and off-specular NR measurements, which took 35\,min and 10\,min, respectively were performed on the FIGARO horizontal sample plane reflectometer\cite{Campbell2011} at the Institut Laue-Langevin using a wavelength ($\lambda$) range from 1.7\,\AA\ to 19\,\AA\ with a relative resolution of $\frac{\Delta\lambda}{\lambda}=4.2$\,\%. The reflectivity was measured at two reflection angles (0.625$^{\circ}$ and 3.2$^{\circ}$) with a relative angular resolution of $\frac{\Delta\theta}{\theta}=3.3$\,\%. The GISANS measurements, which were also performed on FIGARO and took 2\,h each used a wavelength resolution of $\frac{\Delta\lambda}{\lambda}=7.4$\,\% with the same wavelength range as before. The reflection angle was set to $\theta_{i}=0.393^{\circ}$ with a vertical and horizontal angular divergence of $\Delta\theta_{i}=0.02^{\circ}$ and $\Delta\phi_{i}=0.1^{\circ}$, respectively. The detector pixel resolution corresponds to an angular spread of $\Delta\theta_{f}=0.018^{\circ}$ and $\Delta\phi_{f}=0.16^{\circ}$. All resolutions are given as Gaussian equivalent full width at half maximum (FWHM). The solid-liquid sample cell contained about 3\,ml of liquid and was mounted with the liquid on top of the solid so that macroscopic bubbles would drift away from the interface under consideration.\\
The measurements  were taken according to the following procedure.  Firstly specular and off-specuar reflectivities were recorded on the OTS-D$_{2}$O interface by using heavy water as received from \textit{Sigma-Aldrich}, France (99.9 atom \% deuteration). Then the water was exchanged by 9\,ml of ethanol and after 1\,min the ethanol was again exchanged within 100\,s by 12\,ml D$_{2}$O which was saturated with nitrogen by bubbling it with $N_{2}$ for 30\,min at a temperature of 5\,$^{\circ}$C. The sample cell and the ethanol were kept at 45\,$^{\circ}$C throughout the experiment.\par
Fitting of the reflectivity data was accomplished using a slab model with \textit{Motofit} \cite{Motofit}. The error in depletion distance, was determined as follows. The SLD and the thickness of the density depleted layer were allowed to vary simultaneously until the $\chi^{2}$ increased by 5\% as compared to the best fit. This was done because the thickness and SLD of this layer are highly correlated parameters in the data fitting. Then the maximum and minimum values for the depletion distance were calculated within this parameter range. The absolute $\chi^{2}$ values for all NR fits varied between 4 and 5.\\
The GISANS patterns were reduced as follows: In order to accumulate meaningful statistics the data was binned to wavelength bands of $\Delta\lambda/\lambda$\,=\,20\,\%. For the wavelength shown here (5\,\AA) this corresponds to a wavelength range of 4.5-5.5\,\AA\,. This results in a Gaussian equivalent wavelength resolution of $\Delta\lambda/\lambda$\,=\,15\,\% (FWHM). Although other wavelengths were recorded and analyzed as well it turned out that only at 5\,\AA\, wavelength a GISANS signal outside the specular and direct beam region above background could be observed. As can be seen in the measured GISANS patterns the background was very low, on the order of $10^{-6}$ with most of the pixels having 0 counts after 2\,h of acquisition. We therefore did not subtract background from the GISANS images.
\\
The analysis of the GISANS patterns was done by comparing the absolute intensities to simulations performed with the software package BornAgain \cite{BornAgain}. Detector sizes and pixels, incident and reflected angles and divergence as well as wavelength and wavelength resolution were matched to those used in the experiment. The simulations used layer parameters from the specular reflectivity data to form a multilayer structure to which model nanobubbles of oblate shape with different spherical radius $r$, height $h$ and surface coverage $\sigma$ were added at the water-OTS interface. The model used to describe the expected scattering from nanobubbles was the form factor of a truncated oblate spheroid with a scattering length of 0 (air) and the measured contact angle $\theta=(180^{\circ}-93^{\circ})=87^{\circ}$.\\
Structure factor contributions were not included as high nanobubble coverage would be clearly visible by specular reflectivity and would give rise to stronger scattering than that observed, so the simulations focussed on coverages below 25\% (less than half that expected by random sequential adsorption). Polydispersity effects were not included and would correspond to the weighted sum of the images shown in the supporting information. In-plane roughness correlations of the respective interfaces were not included in the simulation as they did not produce scattering stronger than that of the nanobubbles in the range of realistic correlation lengths (<10\,nm).

\section{Results and Discussion}
\begin{figure}[h]
\subfigure[]{\label{fig:NR}\includegraphics[width=7.5 cm]{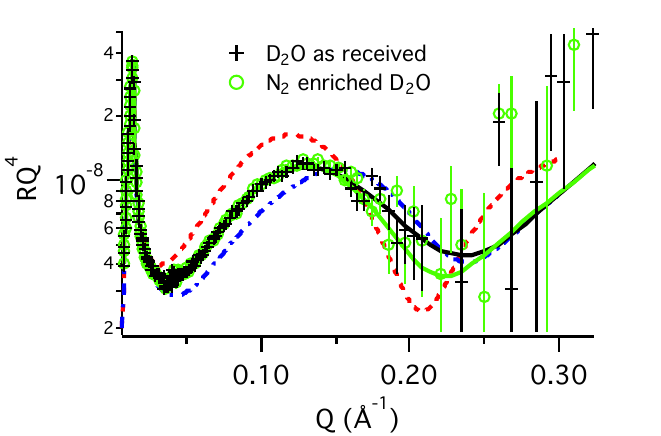}}
\subfigure[]{\label{fig:SLD}\includegraphics[width=7 cm]{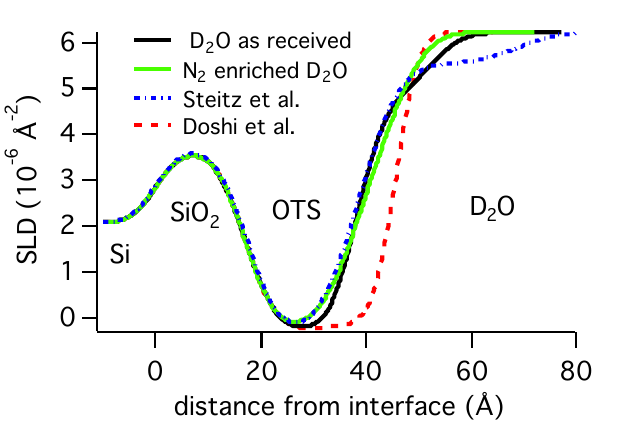}}\\
\caption{\label{fig:NRAll} (a) NR multiplied by q$^{4}$ for the initial D$_{2}$O-OTS interface (black crosses) and the N$_{2}$ enriched water (green circles). The best fits for the two curves are plotted as solid lines in the same color. The dashed-dotted blue line corresponds to a simulation with a depletion layer as observed in\,\cite{Steitz2003}\,, whereas the dashed red line denotes a simulation from a depletion layer as observed in\,\cite{Doshi2005}\,. (b) Scattering length density for the NR fits in the same color code.}
\end{figure}

The thickness and density of the OTS layers were previously determined by X-ray reflectivity on the same batch of samples. The hydrocarbon tail length was thus fixed to 21.3\,\AA\, \cite{Gutfreund2013} corresponding to almost completely stretched chains with a neutron SLD of $-0.4\cdot10^{-6}$\,\AA$^{-2}$ and a porosity of 2\%. The silicon oxide thickness was fixed to 17\,\AA\, including the silane head group of the OTS molecule. The neutron SLD of SiO$_{2}$ was fixed to $3.47\cdot10^{-6}$\,\AA$^{-2}$ and the hydration was determined to be 6\%. The roughnesses of all interfaces were allowed to vary and were between 4 and 6\,\AA.\\  
The specular neutron reflectivity (multiplied by q$^{4}$) of the untreated D$_{2}$O-OTS interface and of the gas enriched water after solvent exchange are plotted in Figure \,\ref{fig:NRAll}. 
In order to fit the data a 5 - 15\,\AA\, thick density depleted layer (50-80\% depletion) sandwiched between the OTS and water had to be assumed leading to a depletion distance of $d_2$\,=\,2.8 - 3.9\,\AA\, in case of the naturally aerated water and a similar range of $d_2$\,=\,1.6 - 3.7\,\AA\, in the case of N$_{2}$ enriched water. These values are only slightly higher than previously mentioned X-ray reflectometry and MD simulation results \cite{Mezger2010} and within the neutron reflectometry experiment of reference \cite{MaccariniSteitz2007}\,. Also the influence of gas enrichment on the depletion distance seems to be negligible, within the error of the measurement, in accordance with earlier X-ray measurements \cite{Mezger2006}. In any case the depletion seems to be of sub-molecular size. The assumption of a thicker density reduced layer analog to that found in reference \cite{Steitz2003} (thickness 35 \AA, roughness 5 \AA, density reduced by $\sim$ 10\,\% leading to $d_2$\,=\,2.6\,\AA\,) impairs the quality of the fit as can be seen in Figure \ref{fig:NR}. The assumption of a zero-density layer as in \cite{Doshi2005} (thickness 7.4\,\AA, roughness 3.4\,\AA, leading to $d_2$\,=\,7.4\,\AA\,) leads to an even worse fit as can be seen in Figure\,\ref{fig:NRAll}.\\
In order to quantify  this result in terms of  the maximum nanobubble surface coverage compatible with these data,  we make some hypothesis on the shape and density of the nanobubble. For simplicity we will limit our analysis to oblate half-sphere shaped objects with a radius $r$ and a height $h$ filled with ambient pressure gas that can be safely assumed to result in zero scattering length density for neutrons. In this case the amount of missing material per unit area can be calculated by multiplying the volume of a oblate half-sphere 2/3$\pi r^{2}h$ with its surface coverage. The resulting value can be directly compared to the depletion distance $d_{2}$ from eq.\,\ref{eq:D2} which also corresponds to the missing material's volume per interface area. By normalizing the nanobubble surface coverage to its surface area at the solid/liquid interface $\pi r^{2}$ we get the relative surface coverage of nanobubbles $\sigma$ in compliance with the depletion distance:
  \begin{equation}
	\sigma=\frac{3d_{2}}{2h}.
	\label{eq:sigma}
\end{equation}
It is evident that, as the height of the nanobubbles decreases, their contribution to the depletion distance will be smaller. We therefore limit the analysis to the minimum nanobubble height detected in literature, which is around 5\,nm \cite{Peng2014} to the best of our knowledge. Hence the maximum surface coverage of oblate half-sphere shaped bubbles compatible with a depletion distance of $d_{2}=3.7$ \AA\, is 11 \%.

\begin{figure}[h]
\includegraphics[width=12 cm]{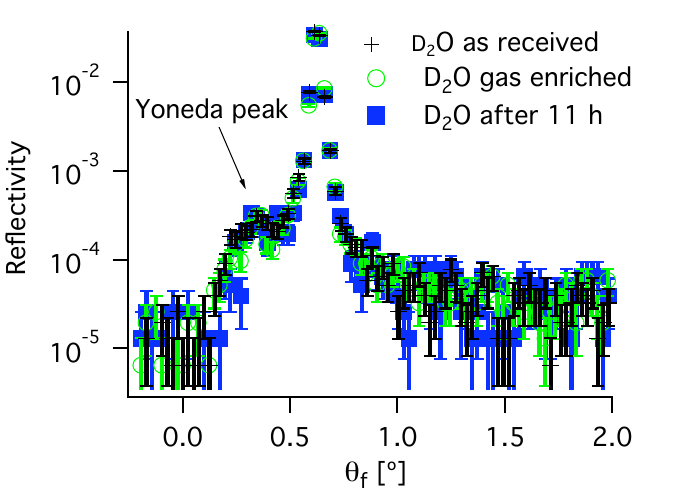}
\caption{Reflected intensity on a logarithmic scale for detector scans at an incident angle of $\theta_{i}=0.625^{\circ}$ and a wavelength of $\lambda=5.5$\,\AA.}
\label{fig:OffSpec} 
\end{figure}
In Fig.\,\ref{fig:OffSpec} we display the off-specular reflectivity curves for the initial interface (black crosses), the gas enriched water immediately after solvent exchange (green circles) and after 11\,h (blue rectangles), respectively. The Yoneda peak intensity originates from an evanescent wave at the solid/liquid interface and is particularly sensitive to in-plane density fluctuations e.g. the presence of nanobubbles adsorbed on the surface. The Yoneda peak appears when the exit angle $\theta_{f}$ matches the critical reflection angle $\theta_{c}$ related to the SLD difference $\Delta N_{b}$ between that of bulk Si and the D$_{2}$O in the vicinity of the solid in the following way:
\begin{equation}
\theta_{c}=\lambda\sqrt{\frac{\Delta N_{b}}{\pi}}.
\label{eq:TotalReflection}
\end{equation}
From fitting a Gaussian to the Yoneda peak in Fig.\,\ref{fig:OffSpec} a critical reflection angle of 0.35$^{\circ}-0.364^{\circ}$ for all three measurements can be derived with no systematic variation between the measurements. By using eq.\,\ref{eq:TotalReflection} a minimum $\Delta N_{b}=3.875\cdot10^{-6}$\AA$^{-2}$ can be derived which leads to a SLD of $5.95\cdot10^{-6}$\AA$^{-2}$ of the adjacent water layer after adding the known Si $N_{b}=2.07\cdot10^{-6}$\AA$^{-2}$. If comparing this value to the bulk water SLD of $6.26\cdot10^{-6}$\AA$^{-2}$ as derived from the specular fits one can estimate a minimum water density of 95\% in the vicinity of the solid. By comparing the volume of a cylinder to a half-sphere of the same radius one can calculate the maximum surface coverage of half-spheroids to be 7.5\% in this case. Moreover, no additional bubbles are introduced within the sensitivity of this measurement.\par
\begin{figure}[h]
\subfigure[]{\label{fig:GISANS}\includegraphics[width=7.75 cm]{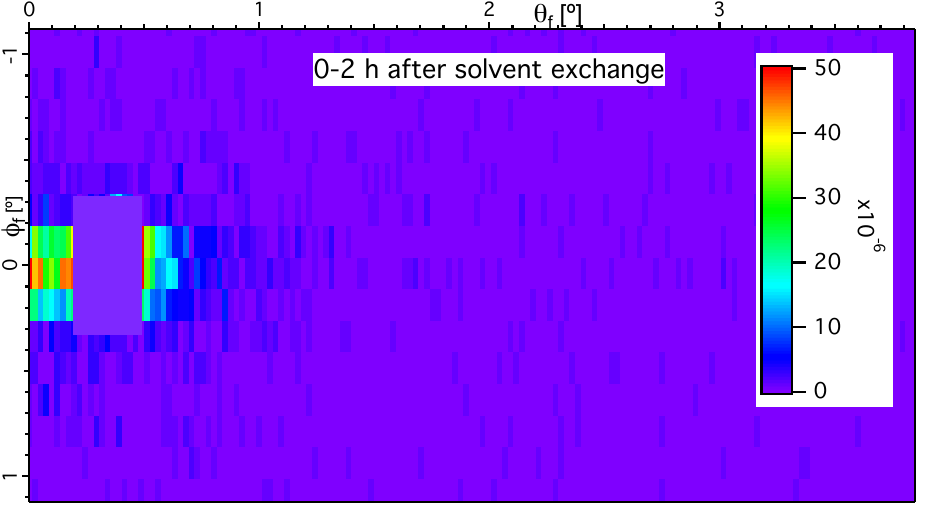}}
\subfigure[]{\label{fig:GISANSSim}\includegraphics[width=8 cm]{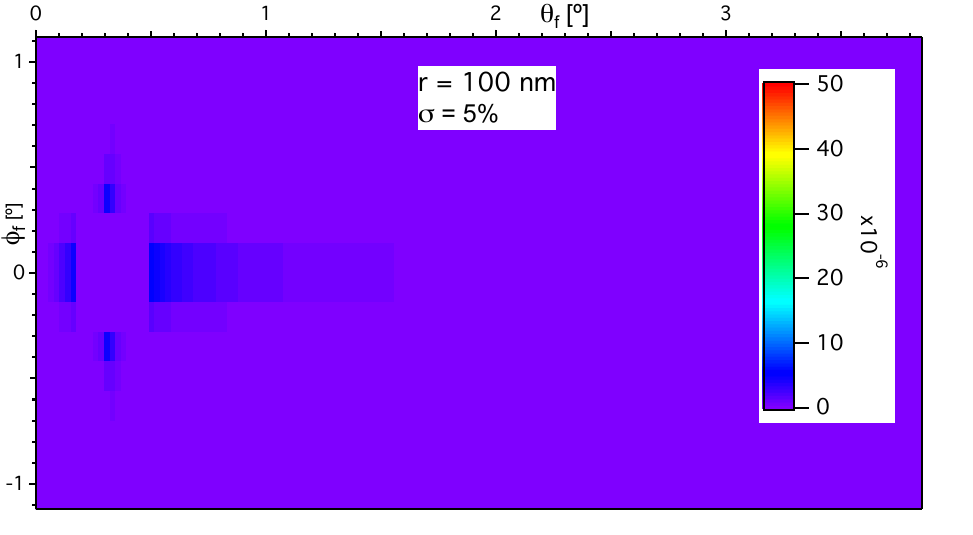}}\\
\subfigure[]{\label{fig:GISANSProj}\includegraphics[width=8 cm]{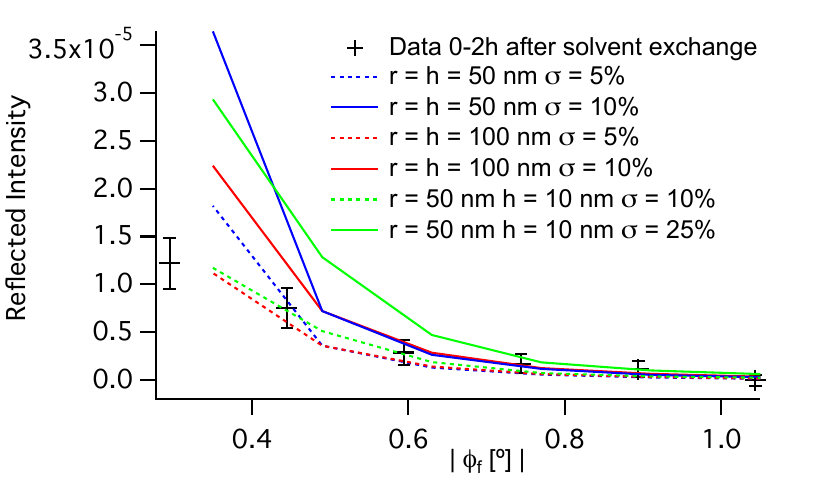}}\\
\caption{\label{fig:GISANSAll} GISANS pattern of the gas enriched water after solvent exchange at 5\,\AA (a) and the corresponding simulation for 100\,nm radius spherical bubbles at 5\% coverage. (b) The specular reflected beam (Intensity $\approx$ 1) is masked. In (c) the projected intensity at the Yoneda peak on the $\phi_{f}$ axis is shown for the data and for various simulations with parameters shown in the legend.}
\end{figure}

In order to check for bubbles in the range of 10\,nm - 400\,nm, we performed GISANS measurements and compared the absolute intensities to simulations performed with the program BornAgain \cite{BornAgain}. A representative pattern after solvent exchange recorded at 5\,\AA\, wavelength and the corresponding simulation at 5\% surface coverage of 100\,nm radius spherical nanobubbles (maximum in the size spectrum of Ref.\,\cite{Zhu2016}) is shown in Fig.\,\ref{fig:GISANSAll}. The measured GISANS pattern does not change significantly even 13 h after solvent exchange (see supporting information). The most prominent feature in the simulated GISANS patterns resulting from nanobubbles is the diffuse scattering along the off-specular axis ($\phi=0$) just above the specular peak (Intensity$\approx1$) which is hidden behind a mask and the lateral lobes extending from the Yoneda scattering (at $\theta_{f}=0.33^{\circ}$). Systematic variation of the nanobubble radius and height in the simulation (see Supplementary Information) predicts that bubbles in the range of 50-100\,nm produce significant intensity in the side-lobes along the Yoneda peak for sufficiently high coverages, which is not as pronounced in the data. For an easier comparison we have plotted an out-of-plane cut along the $\phi_{f}$ axis at the Yoneda postion in Fig.\ref{fig:GISANSProj}. For better statistics a range of $0.26^{\circ}<\theta_{f}<0.36^{\circ}$ was integrated and the neagative and positive $\phi_{f}$ values were binned. Even at a nanobubble concentration as low as 10\% the side-lobe intensity is on the level of $2*10^{-5}$ at an in-plane angle of $\phi_{f}\approx0.4^{\circ}$ for 100\,nm spherical bubbles in the simulation whereas the measured intensity is around $1*10^{-5}$. For oblate nanobubbles the scattering intensity is lower, but even for a radius of 50\,nm (lower limit in Ref.\,\cite{Zhu2016}) and a height of 10\,nm the scattering intensity at 25\% coverage clearly exceeds the measured values. By comparing the measured pattern to the absolute scattering intensities of the simulation at different surface coverages and nanobubble sizes we therefore estimate that the maximum surface coverage compatible with the sensitivity of the technique in this set up is around 10\% for flattened spheroids and even lower for spherical bubbles.

\section{Summary}
In summary, we have shown that with a combination of specular and off-specular neutron reflectometry as well as grazing incidence small angle neutron scattering it is possible to probe a large range of momentum transfers perpendicular and parallel to the surface which correspond to the length-scales present in previous observations of nanobubbles. Nevertheless, we could not find any evidence of a change of nanobubble concentration or depletion layer through the application of the well-established technique of solvent exchange and the use of nitrogen saturated water. Clearly the maximum surface coverage of nanobubbles satisfying the presented results is on a percent level if present at all. We urge the need of quantitative surface averaging techniques for further investigations of nanobubbles. 

\section{Supporting Information Available} 
All simulated GISANS patterns are available in the supporting information. This information is available free of charge via the Internet at http://pubs.acs.org/.

\begin{acknowledgement}
We gratefully acknowledge support from the partnership for soft condensed matter (PSCM) at the ILL and financial support by the Swedish research council VR under contract number A0505501. We further thank Walter van Herck for his help with BornAgain.
\end{acknowledgement}

\bibstyle{achemso}

\begin{mcitethebibliography}{46}
\providecommand*\natexlab[1]{#1}
\providecommand*\mciteSetBstSublistMode[1]{}
\providecommand*\mciteSetBstMaxWidthForm[2]{}
\providecommand*\mciteBstWouldAddEndPuncttrue
  {\def\EndOfBibitem{\unskip.}}
\providecommand*\mciteBstWouldAddEndPunctfalse
  {\let\EndOfBibitem\relax}
\providecommand*\mciteSetBstMidEndSepPunct[3]{}
\providecommand*\mciteSetBstSublistLabelBeginEnd[3]{}
\providecommand*\EndOfBibitem{}
\mciteSetBstSublistMode{f}
\mciteSetBstMaxWidthForm{subitem}{(\alph{mcitesubitemcount})}
\mciteSetBstSublistLabelBeginEnd
  {\mcitemaxwidthsubitemform\space}
  {\relax}
  {\relax}

\bibitem[Ball(2008)]{Ball2008}
Ball,~P. Water as an Active Constituent in Cell Biology. \emph{Chem. Rev.}
  \textbf{2008}, \emph{108}, 74--108, PMID: 18095715\relax
\mciteBstWouldAddEndPuncttrue
\mciteSetBstMidEndSepPunct{\mcitedefaultmidpunct}
{\mcitedefaultendpunct}{\mcitedefaultseppunct}\relax
\EndOfBibitem
\bibitem[Thorsen \latin{et~al.}(2002)Thorsen, Maerkl, and Quake]{Thorsen2002}
Thorsen,~T.; Maerkl,~S.; Quake,~S. Microfluidic large-scale integration.
  \emph{Science} \textbf{2002}, \emph{298}, 580--584\relax
\mciteBstWouldAddEndPuncttrue
\mciteSetBstMidEndSepPunct{\mcitedefaultmidpunct}
{\mcitedefaultendpunct}{\mcitedefaultseppunct}\relax
\EndOfBibitem
\bibitem[Maccarini(2007)]{Maccarini2007}
Maccarini,~M. Water at solid surfaces: A review of selected theoretical aspects
  and experiments on the subject. \emph{Biointerphases} \textbf{2007},
  \emph{2}, MR1--MR15\relax
\mciteBstWouldAddEndPuncttrue
\mciteSetBstMidEndSepPunct{\mcitedefaultmidpunct}
{\mcitedefaultendpunct}{\mcitedefaultseppunct}\relax
\EndOfBibitem
\bibitem[Mamatkulov \latin{et~al.}(2004)Mamatkulov, Khabibullaev, and
  Netz]{Mamatkulov2004}
Mamatkulov,~S.; Khabibullaev,~P.; Netz,~R. Water at hydrophobic substrates:
  Curvature, pressure, and temperature effects. \emph{Langmuir} \textbf{2004},
  \emph{20}, 4756--4763\relax
\mciteBstWouldAddEndPuncttrue
\mciteSetBstMidEndSepPunct{\mcitedefaultmidpunct}
{\mcitedefaultendpunct}{\mcitedefaultseppunct}\relax
\EndOfBibitem
\bibitem[Steitz \latin{et~al.}(2003)Steitz, Gutberlet, Hauss, Kl{\"o}sgen,
  Krastev, Schemmel, Simonsen, and Findenegg]{Steitz2003}
Steitz,~R.; Gutberlet,~T.; Hauss,~T.; Kl{\"o}sgen,~B.; Krastev,~R.;
  Schemmel,~S.; Simonsen,~A.~C.; Findenegg,~G.~H. Nanobubbles and Their
  Precursor Layer at the Interface of Water Against a Hydrophobic Substrate.
  \emph{Langmuir} \textbf{2003}, \emph{19}, 2409--2418\relax
\mciteBstWouldAddEndPuncttrue
\mciteSetBstMidEndSepPunct{\mcitedefaultmidpunct}
{\mcitedefaultendpunct}{\mcitedefaultseppunct}\relax
\EndOfBibitem
\bibitem[Seo and Satija(2006)Seo, and Satija]{Seo2006}
Seo,~Y.-S.; Satija,~S. No intrinsic depletion layer on a polystyrene thin film
  at a water interface. \emph{Langmuir} \textbf{2006}, \emph{22},
  7113--7116\relax
\mciteBstWouldAddEndPuncttrue
\mciteSetBstMidEndSepPunct{\mcitedefaultmidpunct}
{\mcitedefaultendpunct}{\mcitedefaultseppunct}\relax
\EndOfBibitem
\bibitem[Mezger \latin{et~al.}(2010)Mezger, Sedlmeier, Horinek, Reichert,
  Pontoni, and Dosch]{Mezger2010}
Mezger,~M.; Sedlmeier,~F.; Horinek,~D.; Reichert,~H.; Pontoni,~D.; Dosch,~H. On
  the Origin of the Hydrophobic Water Gap: An X-ray Reflectivity and MD
  Simulation Study. \emph{J. Am. Chem. Soc.} \textbf{2010}, \emph{132},
  6735--6741\relax
\mciteBstWouldAddEndPuncttrue
\mciteSetBstMidEndSepPunct{\mcitedefaultmidpunct}
{\mcitedefaultendpunct}{\mcitedefaultseppunct}\relax
\EndOfBibitem
\bibitem[Ocko \latin{et~al.}(2008)Ocko, Dhinojwala, and Daillant]{Ocko2008}
Ocko,~B.~M.; Dhinojwala,~A.; Daillant,~J. Comment on ``{H}ow Water Meets a
  Hydrophobic Surface''. \emph{Phys. Rev. Lett.} \textbf{2008}, \emph{101},
  039601\relax
\mciteBstWouldAddEndPuncttrue
\mciteSetBstMidEndSepPunct{\mcitedefaultmidpunct}
{\mcitedefaultendpunct}{\mcitedefaultseppunct}\relax
\EndOfBibitem
\bibitem[Davis \latin{et~al.}(2012)Davis, Gierszal, Wang, and
  Ben-Amotz]{Davis2012}
Davis,~J.~G.; Gierszal,~K.~P.; Wang,~P.; Ben-Amotz,~D. Water structural
  transformation at molecular hydrophobic interfaces. \emph{Nature}
  \textbf{2012}, \emph{491}, 582--585\relax
\mciteBstWouldAddEndPuncttrue
\mciteSetBstMidEndSepPunct{\mcitedefaultmidpunct}
{\mcitedefaultendpunct}{\mcitedefaultseppunct}\relax
\EndOfBibitem
\bibitem[Schmatko \latin{et~al.}(2005)Schmatko, Hervet, and
  L\'{e}ger]{Schmatko2005}
Schmatko,~T.; Hervet,~H.; L\'{e}ger,~L. Friction and slip at simple fluid-solid
  interfaces: The roles of the molecular shape and the solid-liquid
  interaction. \emph{Phys. Rev. Lett.} \textbf{2005}, \emph{94}, 244501\relax
\mciteBstWouldAddEndPuncttrue
\mciteSetBstMidEndSepPunct{\mcitedefaultmidpunct}
{\mcitedefaultendpunct}{\mcitedefaultseppunct}\relax
\EndOfBibitem
\bibitem[Maccarini \latin{et~al.}(2007)Maccarini, Steitz, Himmelhaus, Fick,
  Tatur, Wolff, Grunze, Janecek, and Netz]{MaccariniSteitz2007}
Maccarini,~M.; Steitz,~R.; Himmelhaus,~M.; Fick,~J.; Tatur,~S.; Wolff,~M.;
  Grunze,~M.; Janecek,~J.; Netz,~R.~R. Density depletion at solid-liquid
  interfaces: A neutron reflectivity study. \emph{Langmuir} \textbf{2007},
  \emph{23}, 598--608\relax
\mciteBstWouldAddEndPuncttrue
\mciteSetBstMidEndSepPunct{\mcitedefaultmidpunct}
{\mcitedefaultendpunct}{\mcitedefaultseppunct}\relax
\EndOfBibitem
\bibitem[Gutfreund \latin{et~al.}(2011)Gutfreund, Wolff, Maccarini, Gerth,
  Ankner, Browning, Halbert, Wacklin, and Zabel]{Gutfreund2011}
Gutfreund,~P.; Wolff,~M.; Maccarini,~M.; Gerth,~S.; Ankner,~J.~F.;
  Browning,~J.~F.; Halbert,~C.~E.; Wacklin,~H.; Zabel,~H. Depletion at
  solid/liquid interfaces: Flowing hexadecane on functionalized surfaces.
  \emph{J. Chem. Phys.} \textbf{2011}, \emph{134}, 064711\relax
\mciteBstWouldAddEndPuncttrue
\mciteSetBstMidEndSepPunct{\mcitedefaultmidpunct}
{\mcitedefaultendpunct}{\mcitedefaultseppunct}\relax
\EndOfBibitem
\bibitem[Gutfreund \latin{et~al.}(2013)Gutfreund, B\"aumchen, Fetzer, van~der
  Grinten, Maccarini, Jacobs, Zabel, and Wolff]{Gutfreund2013}
Gutfreund,~P.; B\"aumchen,~O.; Fetzer,~R.; van~der Grinten,~D.; Maccarini,~M.;
  Jacobs,~K.; Zabel,~H.; Wolff,~M. Solid surface structure affects liquid order
  at the polystyrene--self-assembled-monolayer interface. \emph{Phys. Rev. E}
  \textbf{2013}, \emph{87}, 012306\relax
\mciteBstWouldAddEndPuncttrue
\mciteSetBstMidEndSepPunct{\mcitedefaultmidpunct}
{\mcitedefaultendpunct}{\mcitedefaultseppunct}\relax
\EndOfBibitem
\bibitem[Zhang \latin{et~al.}(2007)Zhang, Khan, and Ducker]{Zhang2007}
Zhang,~X.~H.; Khan,~A.; Ducker,~W.~A. A Nanoscale Gas State. \emph{Phys. Rev.
  Lett.} \textbf{2007}, \emph{98}, 136101\relax
\mciteBstWouldAddEndPuncttrue
\mciteSetBstMidEndSepPunct{\mcitedefaultmidpunct}
{\mcitedefaultendpunct}{\mcitedefaultseppunct}\relax
\EndOfBibitem
\bibitem[Doshi \latin{et~al.}(2005)Doshi, Watkins, Israelachvili, and
  Majewski]{Doshi2005}
Doshi,~D.~A.; Watkins,~E.~B.; Israelachvili,~J.~N.; Majewski,~J. Reduced water
  density at hydrophobic surfaces: {E}ffect of dissolved gases. \emph{PNAS}
  \textbf{2005}, \emph{102}, 9458--9462\relax
\mciteBstWouldAddEndPuncttrue
\mciteSetBstMidEndSepPunct{\mcitedefaultmidpunct}
{\mcitedefaultendpunct}{\mcitedefaultseppunct}\relax
\EndOfBibitem
\bibitem[Mezger \latin{et~al.}(2006)Mezger, Reichert, Sch\"oder, Okasinski,
  Schr\"oder, Dosch, Palms, Ralston, and Honkim\"aki]{Mezger2006}
Mezger,~M.; Reichert,~H.; Sch\"oder,~S.; Okasinski,~J.; Schr\"oder,~H.;
  Dosch,~H.; Palms,~D.; Ralston,~J.; Honkim\"aki,~V. High-resolution in situ
  x-ray study of the hydrophobic gap at the water-octadecyl-trichlorosilane
  interface. \emph{Proc. Natl. Acad. Sci.} \textbf{2006}, \emph{103},
  18401--18404\relax
\mciteBstWouldAddEndPuncttrue
\mciteSetBstMidEndSepPunct{\mcitedefaultmidpunct}
{\mcitedefaultendpunct}{\mcitedefaultseppunct}\relax
\EndOfBibitem
\bibitem[Attard(2003)]{Attard2003}
Attard,~P. \emph{Adv. Coll. Int. Sci.} \textbf{2003}, \emph{104}, 75\relax
\mciteBstWouldAddEndPuncttrue
\mciteSetBstMidEndSepPunct{\mcitedefaultmidpunct}
{\mcitedefaultendpunct}{\mcitedefaultseppunct}\relax
\EndOfBibitem
\bibitem[Brenner and Lohse(2008)Brenner, and Lohse]{Brenner2008}
Brenner,~M.~P.; Lohse,~D. Dynamic Equilibrium Mechanism for Surface Nanobubble
  Stabilization. \emph{Phys. Rev. Lett.} \textbf{2008}, \emph{101},
  214505\relax
\mciteBstWouldAddEndPuncttrue
\mciteSetBstMidEndSepPunct{\mcitedefaultmidpunct}
{\mcitedefaultendpunct}{\mcitedefaultseppunct}\relax
\EndOfBibitem
\bibitem[Colaco(2009)]{Colaco2009}
Colaco,~R. \emph{Surface Science} \textbf{2009}, \emph{603}, 2870\relax
\mciteBstWouldAddEndPuncttrue
\mciteSetBstMidEndSepPunct{\mcitedefaultmidpunct}
{\mcitedefaultendpunct}{\mcitedefaultseppunct}\relax
\EndOfBibitem
\bibitem[Ishida(2000)]{Ishida2000}
Ishida,~N. \emph{Langmuir} \textbf{2000}, \emph{16}, 6377\relax
\mciteBstWouldAddEndPuncttrue
\mciteSetBstMidEndSepPunct{\mcitedefaultmidpunct}
{\mcitedefaultendpunct}{\mcitedefaultseppunct}\relax
\EndOfBibitem
\bibitem[Tyrrell and Attard(2001)Tyrrell, and Attard]{Tyrell2001}
Tyrrell,~J. W.~G.; Attard,~P. Images of Nanobubbles on Hydrophobic Surfaces and
  Their Interactions. \emph{Phys. Rev. Lett.} \textbf{2001}, \emph{87},
  176104\relax
\mciteBstWouldAddEndPuncttrue
\mciteSetBstMidEndSepPunct{\mcitedefaultmidpunct}
{\mcitedefaultendpunct}{\mcitedefaultseppunct}\relax
\EndOfBibitem
\bibitem[Lou \latin{et~al.}(2001)Lou, Gao, Xiao, Li, Li, Zhang, Li, Sun, and
  Hu]{Lou2001}
Lou,~S.; Gao,~J.; Xiao,~X.; Li,~X.; Li,~G.; Zhang,~Y.; Li,~M.; Sun,~J.; Hu,~J.
  Nanobubbles at the liquid/solid interface studied by atomic force microscopy.
  \emph{Chin. Phys.} \textbf{2001}, \emph{10}, S108--S110, Chinese-German
  Workshop on Characterization and Development on Nanosystems, SINO GERMAN CTR
  RES PROMOT, BEIJING, PEOPLES R CHINA, OCT 30-NOV 02, 2000\relax
\mciteBstWouldAddEndPuncttrue
\mciteSetBstMidEndSepPunct{\mcitedefaultmidpunct}
{\mcitedefaultendpunct}{\mcitedefaultseppunct}\relax
\EndOfBibitem
\bibitem[Zhang \latin{et~al.}(2008)Zhang, Quinn, and Ducker]{Zhang2008}
Zhang,~X.~H.; Quinn,~A.; Ducker,~W.~A. Nanobubbles at the Interface between
  Water and a Hydrophobic Solid. \emph{Langmuir} \textbf{2008}, \emph{24},
  4756--4764, PMID: 18366225\relax
\mciteBstWouldAddEndPuncttrue
\mciteSetBstMidEndSepPunct{\mcitedefaultmidpunct}
{\mcitedefaultendpunct}{\mcitedefaultseppunct}\relax
\EndOfBibitem
\bibitem[Karpitschka \latin{et~al.}(2012)Karpitschka, Dietrich, Seddon,
  Zandvliet, Lohse, and Riegler]{Karpitschka2012}
Karpitschka,~S.; Dietrich,~E.; Seddon,~J. R.~T.; Zandvliet,~H. J.~W.;
  Lohse,~D.; Riegler,~H. Nonintrusive Optical Visualization of Surface
  Nanobubbles. \emph{Phys. Rev. Lett.} \textbf{2012}, \emph{109}, 066102\relax
\mciteBstWouldAddEndPuncttrue
\mciteSetBstMidEndSepPunct{\mcitedefaultmidpunct}
{\mcitedefaultendpunct}{\mcitedefaultseppunct}\relax
\EndOfBibitem
\bibitem[Peng \latin{et~al.}(2014)Peng, Birkett, and Nguyen]{Peng2014}
Peng,~H.; Birkett,~G.~R.; Nguyen,~A.~V. Progress on the Surface Nanobubble
  Story: What is in the bubble? Why does it exist? \emph{Adv. Coll. Int. Sci.}
  \textbf{2014}, --\relax
\mciteBstWouldAddEndPuncttrue
\mciteSetBstMidEndSepPunct{\mcitedefaultmidpunct}
{\mcitedefaultendpunct}{\mcitedefaultseppunct}\relax
\EndOfBibitem
\bibitem[Ball(2012)]{Ball2012}
Ball,~P. Nanobubbles are not a Superficial Matter. \emph{ChemPhysChem}
  \textbf{2012}, \emph{13}, 2173--2177\relax
\mciteBstWouldAddEndPuncttrue
\mciteSetBstMidEndSepPunct{\mcitedefaultmidpunct}
{\mcitedefaultendpunct}{\mcitedefaultseppunct}\relax
\EndOfBibitem
\bibitem[Daillant and Gibaud(2009)Daillant, and Gibaud]{Daillant}
Daillant,~J., Gibaud,~A., Eds. \emph{X-ray and Neutron Reflectivity -
  Principles and Applications}; Lect. Notes Phys.; Springer: Heidelberg, 2009;
  Vol. 770\relax
\mciteBstWouldAddEndPuncttrue
\mciteSetBstMidEndSepPunct{\mcitedefaultmidpunct}
{\mcitedefaultendpunct}{\mcitedefaultseppunct}\relax
\EndOfBibitem
\bibitem[Ghosh \latin{et~al.}(2013)Ghosh, Salgin, Pontoni, Reusch, Keil, Vogel,
  Rohwerder, Reichert, and Salditt]{Ghosh2013}
Ghosh,~S.~K.; Salgin,~B.; Pontoni,~D.; Reusch,~T.; Keil,~P.; Vogel,~D.;
  Rohwerder,~M.; Reichert,~H.; Salditt,~T. Structure and Volta Potential of
  Lipid Multilayers: Effect of X-ray Irradiation. \emph{Langmuir}
  \textbf{2013}, \emph{29}, 815--824, PMID: 23231362\relax
\mciteBstWouldAddEndPuncttrue
\mciteSetBstMidEndSepPunct{\mcitedefaultmidpunct}
{\mcitedefaultendpunct}{\mcitedefaultseppunct}\relax
\EndOfBibitem
\bibitem[Squires(1978)]{Squires}
Squires,~G.~L. \emph{Introduction to the Theory of Thermal Neutron Scattering};
  Cambridge University Press: Cambridge, 1978\relax
\mciteBstWouldAddEndPuncttrue
\mciteSetBstMidEndSepPunct{\mcitedefaultmidpunct}
{\mcitedefaultendpunct}{\mcitedefaultseppunct}\relax
\EndOfBibitem
\bibitem[Zabel \latin{et~al.}(2007)Zabel, Theis-Br\"ohl, and
  Toperverg]{Zabel2007}
Zabel,~H.; Theis-Br\"ohl,~K.; Toperverg,~B.~P. In \emph{Novel Techniques for
  Characterizing and Preparing Samples}; Kronm\"uller,~H., Parkin,~S., Eds.;
  Handbook of Magnetism and Advanced Magnetic Materials; John Wiley \& Sons:
  Hoboken, 2007; Vol.~3; Chapter Polarized Neutron Reflectivity and Scattering
  from Magnetic Nanostructures and Spintronic Materials, p 1237\relax
\mciteBstWouldAddEndPuncttrue
\mciteSetBstMidEndSepPunct{\mcitedefaultmidpunct}
{\mcitedefaultendpunct}{\mcitedefaultseppunct}\relax
\EndOfBibitem
\bibitem[Wolff \latin{et~al.}(2009)Wolff, Magerl, and Zabel]{Wolff2009}
Wolff,~M.; Magerl,~A.; Zabel,~H. Crystallization of Soft Crystals.
  \emph{Langmuir} \textbf{2009}, \emph{25}, 64--66\relax
\mciteBstWouldAddEndPuncttrue
\mciteSetBstMidEndSepPunct{\mcitedefaultmidpunct}
{\mcitedefaultendpunct}{\mcitedefaultseppunct}\relax
\EndOfBibitem
\bibitem[Sferrazza \latin{et~al.}(1998)Sferrazza, Heppenstall-Butler, Cubitt,
  Bucknall, Webster, and Jones]{Sferrazza1998}
Sferrazza,~M.; Heppenstall-Butler,~M.; Cubitt,~R.; Bucknall,~D.; Webster,~J.;
  Jones,~R. A.~L. Interfacial Instability Driven by Dispersive Forces: The
  Early Stages of Spinodal Dewetting of a Thin Polymer Film on a Polymer
  Substrate. \emph{Phys. Rev. Lett.} \textbf{1998}, \emph{81}, 5173--5176\relax
\mciteBstWouldAddEndPuncttrue
\mciteSetBstMidEndSepPunct{\mcitedefaultmidpunct}
{\mcitedefaultendpunct}{\mcitedefaultseppunct}\relax
\EndOfBibitem
\bibitem[Salditt(2005)]{Salditt2005}
Salditt,~T. Thermal fluctuations and stability of solid-supported lipid
  membranes. \emph{J. Phys. Condens. Matt.} \textbf{2005}, \emph{17},
  R287\relax
\mciteBstWouldAddEndPuncttrue
\mciteSetBstMidEndSepPunct{\mcitedefaultmidpunct}
{\mcitedefaultendpunct}{\mcitedefaultseppunct}\relax
\EndOfBibitem
\bibitem[Jablin \latin{et~al.}(2011)Jablin, Zhernenkov, Toperverg, Dubey,
  Smith, Vidyasagar, Toomey, Hurd, and Majewski]{Jablin2011}
Jablin,~M.~S.; Zhernenkov,~M.; Toperverg,~B.~P.; Dubey,~M.; Smith,~H.~L.;
  Vidyasagar,~A.; Toomey,~R.; Hurd,~A.~J.; Majewski,~J. In-Plane Correlations
  in a Polymer-Supported Lipid Membrane Measured by Off-Specular Neutron
  Scattering. \emph{Phys. Rev. Lett.} \textbf{2011}, \emph{106}, 138101\relax
\mciteBstWouldAddEndPuncttrue
\mciteSetBstMidEndSepPunct{\mcitedefaultmidpunct}
{\mcitedefaultendpunct}{\mcitedefaultseppunct}\relax
\EndOfBibitem
\bibitem[de~Silva \latin{et~al.}(2009)de~Silva, Martin, Cubitt, and
  Geoghegan]{deSilva2009}
de~Silva,~J.~P.; Martin,~S.~J.; Cubitt,~R.; Geoghegan,~M. Observation of the
  complete rupture of a buried polymer layer by off-specular neutron
  reflectometry. \emph{Europhys. Lett.} \textbf{2009}, \emph{86}, 36005\relax
\mciteBstWouldAddEndPuncttrue
\mciteSetBstMidEndSepPunct{\mcitedefaultmidpunct}
{\mcitedefaultendpunct}{\mcitedefaultseppunct}\relax
\EndOfBibitem
\bibitem[James \latin{et~al.}(2015)James, Higgins, Rees, Geoghegan, Brown,
  Chang, M{\^o}n, Cubitt, Dalgliesh, and Gutfreund]{James2015}
James,~D.; Higgins,~A.~M.; Rees,~P.; Geoghegan,~M.; Brown,~M.~R.; Chang,~S.-S.;
  M{\^o}n,~D.; Cubitt,~R.; Dalgliesh,~R.; Gutfreund,~P. Measurement of
  molecular mixing at a conjugated polymer interface by specular and
  off-specular neutron scattering. \emph{Soft Matter} \textbf{2015}, \emph{11},
  9393--9403\relax
\mciteBstWouldAddEndPuncttrue
\mciteSetBstMidEndSepPunct{\mcitedefaultmidpunct}
{\mcitedefaultendpunct}{\mcitedefaultseppunct}\relax
\EndOfBibitem
\bibitem[Hexemer and M{\"u}ller-Buschbaum(2015)Hexemer, and
  M{\"u}ller-Buschbaum]{MuellerBuschbaum2015}
Hexemer,~A.; M{\"u}ller-Buschbaum,~P. Advanced grazing-incidence techniques for
  modern soft-matter materials analysis. \emph{IUCrJ} \textbf{2015}, \emph{2},
  106--125\relax
\mciteBstWouldAddEndPuncttrue
\mciteSetBstMidEndSepPunct{\mcitedefaultmidpunct}
{\mcitedefaultendpunct}{\mcitedefaultseppunct}\relax
\EndOfBibitem
\bibitem[Metwalli \latin{et~al.}(2011)Metwalli, Moulin, Rauscher, Kaune,
  Ruderer, Buerck, Haese-Seiller, Kampmann, and
  Mueller-Buschbaum]{Metwalli2011}
Metwalli,~E.; Moulin,~J.-F.; Rauscher,~M.; Kaune,~G.; Ruderer,~M.~A.;
  Buerck,~U.~V.; Haese-Seiller,~M.; Kampmann,~R.; Mueller-Buschbaum,~P.
  Structural investigation of thin diblock copolymer films using time-of-flight
  grazing-incidence small-angle neutron scattering. \emph{J. Appl. Cryst.}
  \textbf{2011}, \emph{44}, 84--92\relax
\mciteBstWouldAddEndPuncttrue
\mciteSetBstMidEndSepPunct{\mcitedefaultmidpunct}
{\mcitedefaultendpunct}{\mcitedefaultseppunct}\relax
\EndOfBibitem
\bibitem[Wolff \latin{et~al.}(2014)Wolff, Herbel, Adlmann, Dennison, Liesche,
  Gutfreund, and Rogers]{Wolff2014}
Wolff,~M.; Herbel,~J.; Adlmann,~F.; Dennison,~A. J.~C.; Liesche,~G.;
  Gutfreund,~P.; Rogers,~S. Depth-resolved grazing-incidence time-of-flight
  neutron scattering from a solid{--}liquid interface. \emph{J. Appl. Cryst.}
  \textbf{2014}, \emph{47}, 130--135\relax
\mciteBstWouldAddEndPuncttrue
\mciteSetBstMidEndSepPunct{\mcitedefaultmidpunct}
{\mcitedefaultendpunct}{\mcitedefaultseppunct}\relax
\EndOfBibitem
\bibitem[Durniak \latin{et~al.}(2015)Durniak, Ganeva, Pospelov, Herck, and
  Wuttke]{BornAgain}
Durniak,~C.; Ganeva,~M.; Pospelov,~G.; Herck,~W.~V.; Wuttke,~J. BornAgain -
  Software for simulating and fitting X-ray and neutron small-angle scattering
  at grazing incidence. http://www.bornagainproject.org, 2015\relax
\mciteBstWouldAddEndPuncttrue
\mciteSetBstMidEndSepPunct{\mcitedefaultmidpunct}
{\mcitedefaultendpunct}{\mcitedefaultseppunct}\relax
\EndOfBibitem
\bibitem[Steitz(2004)]{SteitzReport2004}
Steitz,~R. Experimental Report 9-11-1016, 2004\relax
\mciteBstWouldAddEndPuncttrue
\mciteSetBstMidEndSepPunct{\mcitedefaultmidpunct}
{\mcitedefaultendpunct}{\mcitedefaultseppunct}\relax
\EndOfBibitem
\bibitem[Wang and Lieberman(2003)Wang, and Lieberman]{Wang2003}
Wang,~Y.; Lieberman,~M. Growth of ultrasmooth octadecyltrichlorosilane
  self-assembled monolayers on SiO2. \emph{Langmuir} \textbf{2003}, \emph{19},
  1159--1167\relax
\mciteBstWouldAddEndPuncttrue
\mciteSetBstMidEndSepPunct{\mcitedefaultmidpunct}
{\mcitedefaultendpunct}{\mcitedefaultseppunct}\relax
\EndOfBibitem
\bibitem[Campbell \latin{et~al.}(2011)Campbell, Wacklin, Sutton, Cubitt, and
  Fragneto]{Campbell2011}
Campbell,~R.; Wacklin,~H.; Sutton,~I.; Cubitt,~R.; Fragneto,~G. FIGARO: The new
  horizontal neutron reflectometer at the ILL. \emph{Eur. Phys. J. Plus}
  \textbf{2011}, \emph{126}, 1--22\relax
\mciteBstWouldAddEndPuncttrue
\mciteSetBstMidEndSepPunct{\mcitedefaultmidpunct}
{\mcitedefaultendpunct}{\mcitedefaultseppunct}\relax
\EndOfBibitem
\bibitem[Nelson(2006)]{Motofit}
Nelson,~A. Co-refinement of multiple-contrast neutron/X-ray reflectivity data
  using MOTOFIT. \emph{J. Appl. Cryst.} \textbf{2006}, \emph{39},
  273--276\relax
\mciteBstWouldAddEndPuncttrue
\mciteSetBstMidEndSepPunct{\mcitedefaultmidpunct}
{\mcitedefaultendpunct}{\mcitedefaultseppunct}\relax
\EndOfBibitem
\bibitem[Zhu \latin{et~al.}(2016)Zhu, An, Alheshibri, Liu, Terpstra, Liu, and
  Craig]{Zhu2016}
Zhu,~J.; An,~H.; Alheshibri,~M.; Liu,~L.; Terpstra,~P. M.~J.; Liu,~G.;
  Craig,~V. S.~J. Cleaning with Bulk Nanobubbles. \emph{Langmuir}
  \textbf{2016}, Article asap\relax
\mciteBstWouldAddEndPuncttrue
\mciteSetBstMidEndSepPunct{\mcitedefaultmidpunct}
{\mcitedefaultendpunct}{\mcitedefaultseppunct}\relax
\EndOfBibitem
\end{mcitethebibliography}
\providecommand{\latin}[1]{#1}
\providecommand*\mcitethebibliography{\thebibliography}
\csname @ifundefined\endcsname{endmcitethebibliography}
  {\let\endmcitethebibliography\endthebibliography}{}


\end{document}